\documentclass[journal]{IEEEtran}

\usepackage{graphicx}
\usepackage{cite}
\usepackage{subfig}
\usepackage{mathtools, bm}
\usepackage{amssymb, bm}
\newtheorem{proposition}{Proposition}
\newtheorem{definition}{Definition}
\usepackage[mathscr]{euscript}
\usepackage{algorithm}
\usepackage[noend]{algpseudocode}
\usepackage[shortlabels]{enumitem}
\usepackage{xcolor}
\usepackage{hyperref}

\usepackage{amsmath}
\makeatletter
\newcommand*\rel@kern[1]{\kern#1\dimexpr\macc@kerna}
\newcommand*\widebar[1]{%
  \begingroup
  \def\mathaccent##1##2{%
    \rel@kern{0.8}%
    \overline{\rel@kern{-0.8}\macc@nucleus\rel@kern{0.2}}%
    \rel@kern{-0.2}%
  }%
  \macc@depth\@ne
  \let\math@bgroup\@empty \let\math@egroup\macc@set@skewchar
  \mathsurround\z@ \frozen@everymath{\mathgroup\macc@group\relax}%
  \macc@set@skewchar\relax
  \let\mathaccentV\macc@nested@a
  \macc@nested@a\relax111{#1}%
  \endgroup
}

\def\blfootnote{\xdef\@thefnmark{}\@footnotetext}

\makeatother

\DeclareMathOperator*{\argmin}{argmin}
\DeclareMathOperator*{\argmax}{argmax}

\DeclareMathOperator{\vect}{vec}

\begin{document}

\title{Fast Approximation of EEG Forward Problem and Application to Tissue Conductivity Estimation
\thanks{K. Maksymenko is with Athena, Inria Sophia Antipolis, Universit\'e C\^ote d'Azur, France (e-mail: kostiantyn.maksymenko@inria.fr)}
\thanks{M. Clerc is with Athena, Inria Sophia Antipolis, Universit\'e C\^ote d'Azur, France}
\thanks{T. Papadopoulo is with Athena, Inria Sophia Antipolis, Universit\'e C\^ote d'Azur, France}
}
%\subtitle{Do you have a subtitle?\\ If so, write it here}

\author{\IEEEauthorblockN{Kostiantyn~Maksymenko,
Maureen~Clerc, and
Th\'eodore~Papadopoulo}}

\date{Received: date / Accepted: date}
% The correct dates will be entered by the editor
\maketitle

\blfootnote{Copyright (c) 2019 IEEE. Personal use of this material is permitted. However, permission to use this material for any other purposes must be obtained from the IEEE by sending a request to pubs-permissions@ieee.org.}

\begin{abstract}
Bioelectric source analysis in the human brain from scalp electroencephalography (EEG) signals is sensitive to the conductivities of different head tissues. The conductivity of tissues is subject dependent, so non-invasive methods for conductivity estimation are necessary to fine tune EEG models. To do so, the EEG forward problem solution (so-called lead field matrix) must be computed for a large number of conductivity configurations.

Computing a lead field requires a matrix inversion which is computationally intensive for realistic head models. Thus, the required time for computing a large number of lead fields can become impractical. In this work, we propose to approximate the lead field matrix for a set of conductivity configurations, using the exact solution only for a small set of support points in the conductivity space. Our approach accelerates the computation time, while controlling the approximation error.

Our method is tested on simulated and measured EEG data for brain and skull conductivity estimation. This test demonstrates that the approximation does not introduce any bias and runs significantly faster than if exact lead field were to be computed.
\end{abstract}

% Note that keywords are not normally used for peerreview papers.
\IEEEpeerreviewmaketitle

\section{Introduction}
\label{intro}
\IEEEPARstart{T}{he} inverse problem of source reconstruction in electroencephalography (EEG) aims at finding the source distribution that best explains the electric potentials measured noninvasively from electrodes on the scalp surface. A model of the head's electromagnetic transmission plays a central role in accurate source localization. This model is built from a specification of conductivity distribution of the modeled tissue compartments (scalp, skull, cerebrospinal fluid, brain gray and white matter, etc.), which in turn is linked to tissue geometry. Because it is impractical to directly measure head tissue conductivities in vivo for a specific subject, default values are often used. One problem is that the brain-to-skull conductivity ratio reported in the literature varies from 4 to 80 \cite{Rush1968, Goncalves2003, Hoekema2003}. As skull conductivity greatly influences the solution of the forward problem \cite{Vallaghe2008}, localizing brain sources using uncertain conductivity values leads to important errors \cite{AkalinAcar2013, Wang2013, VanUitert2004, Pohlmeier1997}. Taking account of the composite structure of the human skull could improve the accuracy of EEG source analysis \cite{Dannhauer2011}: in such a case, conductivity of each skull tissue should be estimated independently.
 A possible solution is to estimate tissue conductivities and cortical activity simultaneously \cite{Costa2017, Akalin2016, Lew2009, Vallaghe2007}. Doing this requires to solve the EEG forward problem for possibly many different conductivity configurations. The lead field is the linear operator relating brain electrical activity to potentials on EEG electrodes. Computing a lead field requires a matrix inversion which is computationally intensive for realistic head geometry represented with meshes with a large number of vertices. Thus, the required time for computing a large number of solutions for different conductivities quickly becomes impractical.
 
 One way to deal with this problem is to approximate lead fields using a relatively small set of precomputed solutions. The reduced basis method approximates the solution of parametrized PDEs \cite{Hesthaven2016} by reducing the number of degrees of freedom for Galerkin projection based on a set of already computed exact solutions. Our method is inspired by the reduced basis method, but adapted to the particular structure of the EEG forward problem and the nature of its conductivity parameter space. The previous presented method \cite{Costa2017, costa2016eeg} is limited to one unknown conductivity: elements of the lead field matrix are approximated using polynomial interpolation based on a set of precomputed values. With this approach, the complexity of the approximation would increase fast with the number of unknown conductivities.

In this work, we propose a fast lead field approximation method which is robust to head model complexity and to the number of unknown conductivities. 
 
This work is organized as follows. In section \ref{sec:fp}, we recall the EEG forward problem and two numerical methods to solve it. In section \ref{sec:teory}, we present our fast lead field approximation method. Finally, we evaluate the performance of our algorithm on simulated and measured EEG data in section \ref{sec:nr}.

\section{Forward problem}
\label{sec:fp}
\begin{figure}[!t]
	\centering
		\includegraphics[width=2.5in]{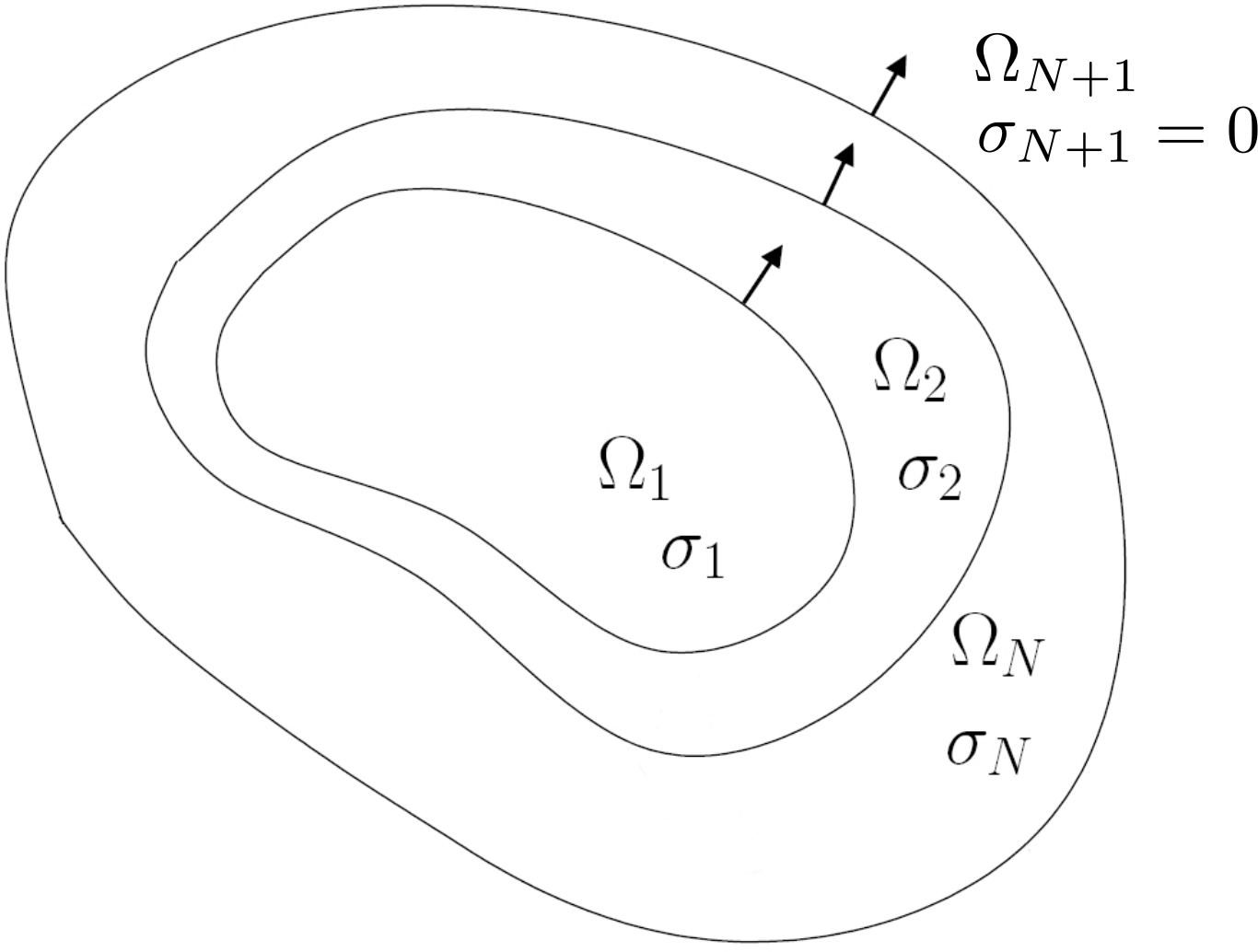}
	\caption{Without loss of generality the head is modeled as a set of regions $\Omega_1,..., \Omega_{N}$ with constant isotropic conductivities $\sigma_1,...,\sigma_{N}$. Adapted from \cite{Kybic2005}.}
	\label{fig:layers}
\end{figure}

The useful frequency range for electrophysiological signals in MEG and EEG is typically below 1 kHz, and most studies deal with frequencies between 0.1 and 100 Hz. Consequently, the physics of MEG and EEG can be
described by the quasi-static approximation of Maxwell's equations \cite{Hamalainen1993}. In a conductive environment, it yields the fundamental Poisson equation with boundary condition :
\begin{equation} 
  \label{eq.maxwell}
  \begin{cases}
  \nabla \cdot (\sigma \nabla V) = \nabla \cdot J ~~\text{   in } \Omega \\
  \sigma \dfrac{\partial V}{\partial \mathbf{n}} = \sigma \nabla V \cdot \mathbf{n} = 0 ~~\text{ on } \partial \Omega ~,
  \end{cases}
\end{equation}
where $\Omega \subset \mathbb{R}^3$ is the head domain, $\partial \Omega$ is its boundary with outward pointing normal vector $\mathbf{n}$, $\sigma [S / m]$ is the conductivity, $J [A/ m^2]$ is the primary current source density, supposed to be known in the forward problem, and $V ~[V]$ is the unknown electric potential. It is usually assumed that the head is composed of several subdomains $\Omega_n$ with different electrical conductivities $\sigma_n$. We assume a piecewise constant conductivity model (Fig. \ref{fig:layers}). Conductivity is supposed isotropic and constant in each domain $\Omega_{i}$. The different tissue conductivities are denoted $\boldsymbol{\sigma} = (\sigma_1,\cdots,\sigma_{N})$.

\subsection{Numerical approximation of the forward problem.}

Equation (\ref{eq.maxwell}) has no analytic solution for realistic (nonspherical) head geometries, but numerical methods can be used to approximate its solution. We will focus on the finite element method (FEM) \cite{Wolters2004,Vallaghe2010} and the boundary element method (BEM)\cite{Sarvas1987, Kybic2005}, which are both based on the Galerkin approximation. The method presented in this paper applies both to FEM and BEM but will be illustrated on the symmetric BEM \cite{Kybic2005}.

For both FEM and BEM approaches, the initial continuous problem (\ref{eq.maxwell}) is discretized to a linear system of type:

\begin{equation}
\label{eq:system}
H_{\boldsymbol{\sigma}}\mathbf{v} = \mathbf{d}(J). 
\end{equation}
Let $N_V$ be the number of unknowns in the head model. Vector $\mathbf{v} \in \mathbb{R}^{N_V}$ represents the unknowns of the model, which are values of the potential on the mesh nodes for FEM and standard BEM, and potentials on mesh nodes and their normal derivatives on the mesh triangles for symmetric BEM. The $N_V \times N_V$ matrix $H_{\boldsymbol{\sigma}}$, called "head matrix", can be computed once head geometry, conductivity values and finite element basis functions are fixed. Vector $\mathbf{d}(J)$ depends on the source configuration. Notice that $H_{\boldsymbol{\sigma}}$ depends on conductivity. An important property of matrix $H_{\boldsymbol{\sigma}}$ is that, under the assumption that the head is composed of several subdomains with constant conductivity, it can be represented as a linear combination of \textit{conductivity-independent} matrices:
\begin{equation}
\label{eq:sum_h}
H_{\boldsymbol{\sigma}} = \sum_{i=1}^{N_H} \gamma_i(\boldsymbol{\sigma}) \widebar{H}_i,
\end{equation}
where $\gamma_i(\boldsymbol{\sigma})$ are scalar functions, $N_H$ represents the number of conductivity-independent components $\widebar{H}_i$. For FEM, $\gamma_i(\boldsymbol{\sigma}) = \sigma_i$ (can be seen in Section 3 of \cite{Wolters2004}). For symmetric BEM, multipliers $\gamma_i(\boldsymbol{\sigma})$ have more complex structure, for example $\{-\sigma_i, ~\sigma^{-1}_i, ~\sigma_i + \sigma_j, ~\sigma^{-1}_i + \sigma^{-1}_j, ~\cdots\}$ (equation (28) in \cite{Kybic2005}). As an example, for a three-layer nested head model with brain, skull and scalp, $N_H = 3$ for FEM and $N_H = 7$ for BEM.

\subsection{Source model}

The most common source model  to represent electrical activity in the brain is a "current dipole". It represents an oriented source of current located at a single position $\mathbf{r}_0$, with dipolar moment $\mathbf{q}$, and it is denoted by $J(\mathbf{r}) = \mathbf{q}\delta(\mathbf{r}-\mathbf{r}_0)$, where $\delta(\cdot)$ is a Dirac distribution.

The source space can be seen as a finite set of $N_S$ dipoles with known positions. Without loss of generality, we can also assume that dipole orientations are known. Let $x_i$ represent the amplitude of the $i$-th source and $\widebar{J}_i$ the $i$-th source moment with unit amplitude:
\[
J(\mathbf{r}) = \sum_{i=1}^{N_S} x_i \widebar{J}_i(\mathbf{r}).
\]
The source term $\mathbf{d}$ in (\ref{eq:system}) is linear with respect to $J$, therefore:
\[
\mathbf{d}(J) = \sum_{i=1}^{N_S} x_i \mathbf{d}(\widebar{J}_i(\mathbf{r})) = \sum_{i=1}^{N_S} x_i \mathbf{d}_i = D\mathbf{x}~,
\]
where the $i$-th column of the $N_V \times N_S$ matrix $D$ corresponds to the $i$-th unit source and $\mathbf{x} = (x_1, \cdots, x_{N_S})$ represents the source amplitudes. Let us notice that in the case of FEM, matrix $D$ does not depend on conductivities. For BEM, it does, so we will note it $D_{\boldsymbol{\sigma}}$ for generality. Moreover, $D_{\boldsymbol{\sigma}}$ can be represented as follows:
\begin{equation}
\label{eq:sum_d}
D_{\boldsymbol{\sigma}} = \sum_{i=1}^{N_D} \lambda_i(\boldsymbol{\sigma}) \widebar{D}_i ~,
\end{equation}
where matrices $\widebar{D}_i$ are independent of $\boldsymbol{\sigma}$ and $\lambda_i(\boldsymbol{\sigma})$ are scalars. In the case of symmetric BEM $\lambda_i(\boldsymbol{\sigma})$ are $\{1, \sigma_i, \cdots\}$ (equation (28) in \cite{Kybic2005}).

The linear system to solve then becomes:
\begin{equation}
\label{eq:matrix_form}
H_{\boldsymbol{\sigma}}\mathbf{v} = D_{\boldsymbol{\sigma}}\mathbf{x}.
\end{equation}

\subsection{Lead field matrix}
In the context of EEG, the \textit{lead field} is a linear operator that maps source activation to potentials at sensor locations:
$$\mathbf{v}_{eeg} = L\mathbf{x}.$$

Each column of the lead field matrix $L$ represents the contribution of the corresponding unit norm source on EEG electrodes. Computing $\mathbf{v}_{eeg}$ requires applying to $\mathbf{v}$ a matrix $S$ which selects or interpolates potentials at electrode positions:  $\mathbf{v}_{eeg} = S \mathbf{v}$. We denote by $N_E$ the number of EEG electrodes.

Using (\ref{eq:matrix_form}) and the selection matrix $S$:
\[
\mathbf{v}_{eeg} = SH_{\boldsymbol{\sigma}}^{-1}D_{\boldsymbol{\sigma}}\mathbf{x}.
\]

Let us remark that since the electric potential is only defined up to a constant, the head matrix $H_{\boldsymbol{\sigma}}$ is not full rank and has a one-dimensional kernel (consant vector). So the inverse notation $H_{\boldsymbol{\sigma}}^{-1}$ actually implies a deflation \cite{Chan1984} which is usually applied in this type of situation.

The $N_E \times N_S$ lead field matrix can thus be computed as:
\begin{equation}
\label{eq:leadfield}
	L = SH_{\boldsymbol{\sigma}}^{-1}D_{\boldsymbol{\sigma}}.
\end{equation}

\section{Fast lead field approximation method}
\label{sec:teory}

\subsection{Motivation}

As already mentioned in introduction, in this work we are interested in computing the EEG forward problem for possibly many different conductivity configurations (with potential application to tissue conductivity estimation). The numerical complexity of lead field computation is essentially due to two operations:

\begin{itemize}
    \item computation of matrices $\widebar{H}_i, i=1\ldots N_H$ and $\widebar{D}_i, i=1\ldots N_D$,
    
    \item inversion of head matrix $H_{\sigma}$.
    
\end{itemize}

Matrices $\widebar{H}_i$ and $\widebar{D}_i$ do not depend on conductivity, therefore they need only to be computed once. The inverse of $H_{\sigma}$, however, depends on $\sigma$ (\ref{eq:sum_h}) and, in general, has to be recomputed for each conductivity configuration of interest. For realistic head models implying a large number of unknowns, the size of matrix $H_{\sigma}$ is relatively large. As a result, in the case of a large set of conductivity configurations, the time required for inversion of head matrices -- and therefore for computing lead fields -- becomes prohibitive. In this context, it can be worthwhile to compute not the \textit{exact} lead fields for all needed conductivity configurations, but an \textit{approximation} thereof.

In this section, we propose a lead field approximation method based on the following ideas. First of all, we assume that the manifold of parametrized solutions of EEG forward problem can be approximated by low-dimensional linear subspace. We choose a domain of interest in conductivity space, spanning a range of values for the head tissue conductivities. We then select \textit{support points} within this domain, for which \textit{exact} forward solutions will be computed. Based on these exact solutions a lead field \textit{approximation} can be computed for any other conductivity configuration.

\begin{figure}[!t]
	\centering
		\includegraphics[width=0.4\textwidth]{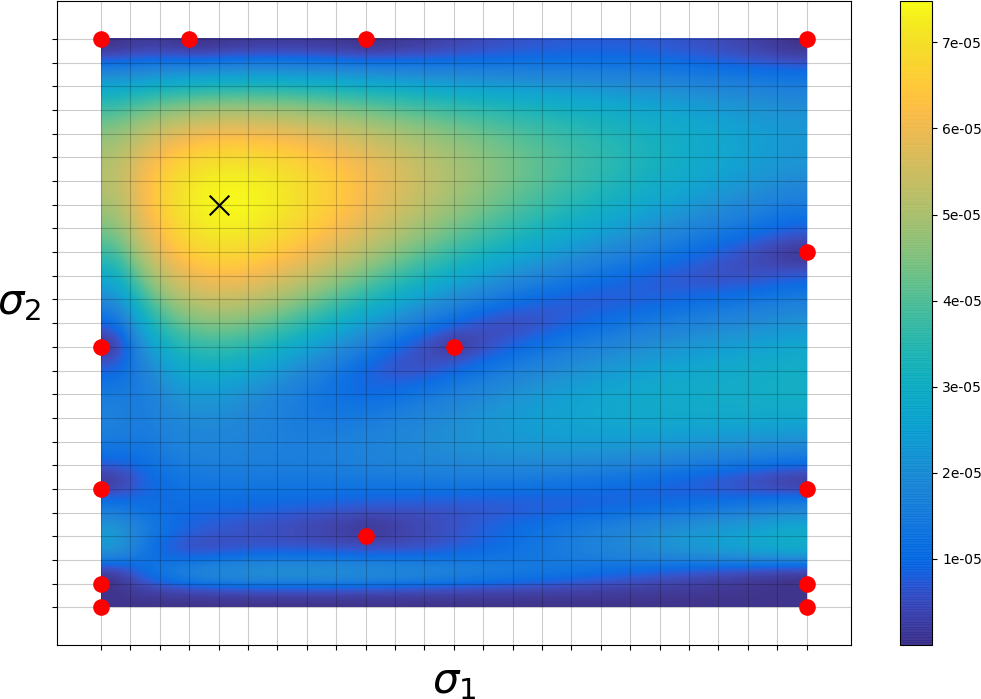}
	\caption{Schematic representation of a greedy support points selection. The domain of interest, here a two dimensional conductivity region, is sampled with a mesh grid. Red points correspond to current support points. The color represents upper bound error, computed at grid points (with sub-grid interpolation only for visual purposes).  The grid point which maximises this error (black cross) is selected as the next support point. Let us notice that even if support points are selected from a finite sampling of the domain of interest, the lead field can be approximated for any conductivity configuration in the domain.}
	\label{fig:basis_points}
\end{figure}

Support points are selected via a greedy algorithm. Similarly to the greedy basis generation used in reduced basis methods for parametrized partial differential equations \cite{Hesthaven2016}, our approach is an iterative procedure where each iteration adds one new support point. Each new support point is the one which maximizes the upper bound approximation error, over a sampling of the domain of interest. Fig. \ref{fig:basis_points} shows a schematic representation of this idea.

\textit{Remark:} We make the assumption that the lead field as a function of conductivity
is sufficiently regular so that the continuous conductivity domain can
be explored using a sufficiently dense sampling.

In subsection \ref{sec:formal}, we formally define the approximation problem and upper bound for the approximation error. In subsection \ref{sec:approx}, we propose a particular choice for this error which is appropriate for the lead field approximation problem. Finally in subsection \ref{sec:algo}, we describe the greedy algorithm for adding support points. 

\subsection{Lead field approximation strategy}
\label{sec:formal}
To approximate the true lead field matrix $L(\boldsymbol{\sigma})$, we introduce coefficients $\boldsymbol{\alpha} \in \mathbb{R}^n$ where $n$ is the number of support points. The lead field approximation is a matrix denoted by $L_n(\boldsymbol{\alpha},\boldsymbol{\sigma})$. We will see later that parameters $\boldsymbol{\alpha}$ represent coefficients of linear approximation. We consider the following relative approximation error, using the Frobenius matrix norm $\|L\|_F = (\sum l_{ij}^2)^{\frac{1}{2}}$:
\begin{equation}
\label{eq:error}
E(L(\boldsymbol{\sigma}),L_n(\boldsymbol{\alpha}, \boldsymbol{\sigma})) = \frac{\left\lVert  L(\boldsymbol{\sigma}) - L_n(\boldsymbol{\alpha}, \boldsymbol{\sigma}) \right\rVert_F}{\left\lVert  L(\boldsymbol{\sigma}) \right\rVert_F }.
\end{equation}

To control this error without knowing $L(\boldsymbol{\sigma})$, we introduce the definition of an upper bound for $E(L,L_n)$. The idea is to bound $E$ with an error function $E_n$, which can be easily computed, converges to zero when the number of support points $n$ increases, and thus ensures the convergence of $E$ to zero.

\begin{definition}
\label{def:upper_bound}
For $n \in \mathbb{N}$, function $E_n(\boldsymbol{\alpha},\boldsymbol{\sigma})$ is an \textit{upper bound approximation} of $E(L(\boldsymbol{\sigma}),L_n(\boldsymbol{\alpha}, \boldsymbol{\sigma}))$ if: 
\begin{multline*}
~\exists C>0 \mbox{ such that }\forall \boldsymbol{\alpha} \in \mathbb{R}^{n},\\
~E(L(\boldsymbol{\sigma}),L_n(\boldsymbol{\alpha}, \boldsymbol{\sigma})) \leq C \cdot E_n(\boldsymbol{\alpha},\boldsymbol{\sigma}).
\end{multline*}
\end{definition}

Let us remark that $C>0$ is a constant which does not depend on $\boldsymbol{\alpha}$ but may depend on $\boldsymbol{\sigma}$.

As $E$ is non-negative, it can be directly derived from definition \ref{def:upper_bound} that if $E_n\underset{n \to \infty}{\longrightarrow} 0$ then $E(L, L_n)\underset{n \to \infty}{\longrightarrow} 0 ~$.

\begin{definition}
\label{def:opt}
The sequence of matrices $\{L_n(\boldsymbol{\alpha^*}_n(\boldsymbol{\sigma}),\boldsymbol{\sigma}), n \in \mathbb{N}\}$ is a \textit{valid approximation} of $L(\boldsymbol{\sigma})$ with \textit{optimal approximation parameters} $\{\boldsymbol{\alpha^*}_n(\boldsymbol{\sigma}), n \in \mathbb{N} \}$ if both following conditions hold:
\begin{enumerate}[(i)]
\item $\forall n, ~\boldsymbol{\alpha^*}_n(\boldsymbol{\sigma}) = \argmin_{\boldsymbol{\alpha}} E_n(\boldsymbol{\alpha},\boldsymbol{\sigma})$ ,

\item $E_n(\boldsymbol{\alpha^*}_n(\boldsymbol{\sigma}),\boldsymbol{\sigma})\underset{n\to \infty}{\longrightarrow} 0$.
\end{enumerate}
\end{definition}
The following sections will propose explicit formulations for the upper bound error and for coefficients $\boldsymbol{\alpha}$ as well as algorithms to select the support points.

\subsection{Choice of the upper bound approximation}\label{sec:approx}

As mentioned previously, the objective of the approximation is to circumvent the inversion of matrix $H_{\boldsymbol{\sigma}}$. One possible way to parametrize the approximation matrix $L_n(\boldsymbol{\alpha},\boldsymbol{\sigma})$ is to approximate the inverse $H_{\boldsymbol{\sigma}}^{-1}$ as a linear combination of precomputed inversions $H_{\boldsymbol{\sigma}_i}^{-1}$ at $n$ support points:

\begin{equation}
\label{eq:hinv}
H_{\boldsymbol{\sigma}}^{-1} \approx \sum_{i=1}^{n}\alpha_iH_{\boldsymbol{\sigma}_i}^{-1} ~.
\end{equation}

Based on (\ref{eq:leadfield}), the proposed lead field approximation takes the form:
\[
L_n(\boldsymbol{\alpha},\boldsymbol{\sigma}) = \sum_{i=1}^{n}\alpha_i SH_{\boldsymbol{\sigma}_i}^{-1}D_{\boldsymbol{\sigma}} ~.
\]

Using the linear decomposition of $D_{\boldsymbol{\sigma}}$ in (\ref{eq:sum_d}), we represent the lead field approximation as a linear combination of matrices which are independent of $\boldsymbol{\sigma}$:

\begin{multline}
\label{eq:lf-approx}
L_n(\boldsymbol{\alpha},\boldsymbol{\sigma}) = \sum_{i=1}^{n}\sum_{j=1}^{N_D}\alpha_i \lambda_j(\boldsymbol{\sigma}) SH_{\boldsymbol{\sigma}_i}^{-1} \widebar{D}_j =\\
\sum_{i=1}^{n}\sum_{j=1}^{N_D}\beta_{ij}(\boldsymbol{\sigma},\boldsymbol{\alpha})\widebar{L}_{ij} ~,
\end{multline}
where $\widebar{L}_{ij} = SH_{\boldsymbol{\sigma}_i}^{-1}\widebar{D}_j$ are matrices which do not depend on conductivity and thus can be precomputed, and $\boldsymbol{\beta}_{ij}(\boldsymbol{\sigma},\boldsymbol{\alpha}) = \alpha_i \lambda_j(\boldsymbol{\sigma})$ are conductivity-dependent scalars.  

As  $\lambda_j$ are known (\ref{eq:sum_d}), the question is how to compute optimal coefficients $\boldsymbol{\alpha}$. We will base our approach on the property of an inverse matrix: $H_{\boldsymbol{\sigma}}^{-1} \cdot H_{\boldsymbol{\sigma}} = I$, where $I$ is the identity matrix. In the context of the EEG forward problem, it is more relevant to consider matrix $SH^{-1}_{\boldsymbol{\sigma}}$, which has smaller dimensions than $H^{-1}_{\boldsymbol{\sigma}}$ and which contains only the relevant part of $H^{-1}_{\boldsymbol{\sigma}}$. Having $SH_{\boldsymbol{\sigma}}^{-1} \cdot H_{\boldsymbol{\sigma}} = S$ and (\ref{eq:hinv}), we could seek $\boldsymbol{\alpha}$ by minimizing the expression:
\begin{equation}
\label{eq:prop1}
	{\boldsymbol{\alpha}}^*_n(\boldsymbol{\sigma}) = \argmin_{\boldsymbol{\alpha}}\left\lVert S - \sum_{i=1}^n \alpha_i SH_{\boldsymbol{\sigma}_i}^{-1}H_{\boldsymbol{\sigma}}\right\rVert_F ~.
\end{equation}

Proposition \ref{prop:1} shows that this choice of $\boldsymbol{\alpha}$ leads to an upper bound approximation in the sense of Definition \ref{def:upper_bound}. 

\begin{proposition}
\label{prop:1}
If matrices $\{SH_{\boldsymbol{\sigma}_i}^{-1}H_{\boldsymbol{\sigma}}\}_{i=1}^n$ are linearly independent, then 

\begin{equation*}
E_n(\boldsymbol{\alpha} ,\boldsymbol{\sigma}) =\left\lVert S - \sum_{i=1}^n \alpha_i SH_{\boldsymbol{\sigma}_i}^{-1}H_{\boldsymbol{\sigma}}\right\rVert_F
\end{equation*}
is an upper bound approximation of error $E(L,L_n)$ (\ref{eq:error}) and $\boldsymbol{\alpha}^*_n(\boldsymbol{\sigma})$ is an optimal approximation parameter.
\end{proposition}
\begin{IEEEproof}
See Appendix \ref{app:a}.
\end{IEEEproof}

Linear dependence of \textit{basis matrices} $\{SH_{\boldsymbol{\sigma}_i}^{-1}H_{\boldsymbol{\sigma}}\}_{i=1}^n$, would mean that lead fields corresponding to some support points could be exactly represented as a linear combination of lead fields at other support points. So the linear independence condition of Proposition \ref{prop:1} can be handled by  properly selecting support points. This will be discussed in detail in the next section.

We now express problem (\ref{eq:prop1}) as a simple least squares problem. Using the linear decomposition of $H_{\boldsymbol{\sigma}}$ as a sum of conductivity-independent matrices (\ref{eq:sum_h}), we get:

\[
	SH_{\boldsymbol{\sigma}_i}^{-1}H_{\boldsymbol{\sigma}} = \sum_{j=1}^{N_H}  \gamma_j(\boldsymbol{\sigma}) SH_{\boldsymbol{\sigma}_i}^{-1}\widebar{H}_j ~.
\]

We can then vectorize matrices $SH_{\boldsymbol{\sigma}_i}^{-1}\widebar{H}_j$ and assemble them as columns of a new matrix $K$. Let $K_{., (i-1)N_H + j} = \vect(SH_{\boldsymbol{\sigma}_i}^{-1}\widebar{H}_j)$ denote the $((i-1)N_H + j)$-th column of $K$. The matrix $K$ will thus have $n \cdot N_H$ columns and $N_E \cdot N_V$ rows. 

Let us also denote $\Gamma_{\boldsymbol{\sigma}} = I \otimes \boldsymbol{\gamma}(\boldsymbol{\sigma})$, where $\boldsymbol{\gamma}(\boldsymbol{\sigma}) = (\gamma_1(\boldsymbol{\sigma}),\cdots, \gamma_{N_H}(\boldsymbol{\sigma}))$ and $I$ is an identity matrix of size $n$ (number of support points):
$$ \Gamma_{\boldsymbol{\sigma}} = 
\begin{bmatrix}
    \gamma_1(\boldsymbol{\sigma}) & 0 & \dots  & 0 \\
    \vdots & \vdots  & \ddots  & \vdots \\
    \gamma_{N_H}(\boldsymbol{\sigma}) & 0 & \dots & 0 \\
    0  & \gamma_1(\boldsymbol{\sigma}) & \dots  & 0 \\
    \vdots  & \vdots & \ddots  & \vdots \\
    0  & \gamma_{N_H}(\boldsymbol{\sigma}) & \dots  & 0 \\
    \vdots  & \vdots & \ddots  & \vdots \\
    0  & 0  &\dots  & \gamma_1(\boldsymbol{\sigma}) \\
    \vdots  & \vdots  & \ddots  & \vdots \\
    0  & 0  & \dots  & \gamma_{N_H}(\boldsymbol{\sigma})
\end{bmatrix}
$$

Using the introduced notations, we can show that $K\Gamma_{\boldsymbol{\sigma}}$ represents a linear projection basis and that the solution $\boldsymbol{\alpha}^*_n(\boldsymbol{\sigma})$ of problem (\ref{eq:prop1}) is given by the solution of the system:
\begin{equation}
\label{eq:opt2}
(\Gamma^T_{\boldsymbol{\sigma}} K^T K \Gamma_{\boldsymbol{\sigma}}) \boldsymbol{\alpha}^*_n(\boldsymbol{\sigma}) = \Gamma^T_{\boldsymbol{\sigma}}K^T\vect(S) ~.
\end{equation} 
See Appendix \ref{app:b} for derivation.

Some important remarks must be noted:
\begin{enumerate}
\item The size of matrix $\Gamma_{\boldsymbol{\sigma}}^T K^T K \Gamma_{\boldsymbol{\sigma}}$ is $n \times n$. The complexity of problem (\ref{eq:opt2}) therefore does not depend on the size of the head model but only on the number of support points. 
\item Matrices $K^TK$ and $K^T \vect(S)$ are independent of $\boldsymbol{\sigma}$. Moreover, adding a new support point amounts to adding new elements to these matrices - they do not have to be fully recomputed.
\end{enumerate}

To summarize, given a set of $n$ support points, we use (\ref{eq:opt2}) to compute the optimal approximation parameter $\boldsymbol{\alpha}^*_n(\boldsymbol{\sigma})$ and then use (\ref{eq:lf-approx}) to compute lead field approximation $L_n(\boldsymbol{\alpha}^*_n(\boldsymbol{\sigma}),\boldsymbol{\sigma})$.

\subsection{Selection of support points}
\label{sec:algo}
The method introduced in subsection \ref{sec:approx} supposes that a set of support points is given allowing to precompute matrices $K^TK$, $K^T\vect(S)$ and $\widebar{L}_{ij}$. This section answers the question of how to select them. 

% First, let us assume that the upper bound error $E_n(\boldsymbol{\sigma})$ is continuous on the domain of interest. 
The idea is to start with a small number of support points and to add new points iteratively. At each iteration, we select a conductivity which maximizes the upper bound error on a discretization of the conductivity domain of interest. The procedure is summarized in Algorithm \ref{alg:bp}.

\begin{algorithm}
\caption{Support points selection}\label{alg:bp}
\textbf{Input}: 
Sampling $\Sigma = \{\boldsymbol{\sigma}_i\}_{i=1}^M$ of the domain of interest ;
set $\text{Supp}$ of initial support points.
\begin{algorithmic}[1]
\State $n \leftarrow |\text{Supp}|$.
\State Compute matrices $K^TK$ and $K^T\vect(S)$.
\Repeat
\For{each $\boldsymbol{\sigma}$ in $\Sigma$}
\State Compute $\boldsymbol{\alpha}_n^*(\boldsymbol{\sigma})$ with (\ref{eq:opt2}).
\State Compute $E_n(\boldsymbol{\alpha_n^*}(\boldsymbol{\sigma}),\boldsymbol{\sigma})$.
\EndFor
\State $\text{Supp} = \text{Supp} \cup \argmax_{\boldsymbol{\sigma} \in \Sigma} \{E_n(\boldsymbol{\alpha_n^*}(\boldsymbol{\sigma}),\boldsymbol{\sigma})$
\State Update matrices $K^TK$ and $K^T\vect(S)$.
\State $n \leftarrow n + 1$
\Until{convergence}
\end{algorithmic}
\textbf{Output} :
Matrices $K^TK$ and $K^T\vect(S)$. 
\end{algorithm}

\begin{itemize}
    \item In this paper, the discretization of the domain of interest was uniform, but this choice is arbitrary: the sampling can be chosen according to the application.
    
    \item In this work we initialized the basis with the corners of the domain of interest (e. g. $4$ support points for the $2D$ domain). This choice is arbitrary, and Algorithm \ref{alg:bp} can be initialized with for instance a single support point.
    
    \item There are several ways to define a convergence criterion. Even though the real approximation error is unknown, it seems reasonable to control the maximum value of the upper bound error, noted $E^*_n$ (stop when $E^*_n < \epsilon$), or its decreasing rate (stop when $|E^*_n - E^*_{n-1}| < \epsilon$).
\end{itemize}

Algorithm \ref{alg:bp} has interesting properties which make it more efficient than a random selection of support points.

First, the valid upper bound error is controlled, in a strong sense: 
\[
\forall \boldsymbol{\sigma} ~E_{n+1}(\boldsymbol{\alpha^*}_{n+1}(\boldsymbol{\sigma}),\boldsymbol{\sigma}) \leq E_{n}(\boldsymbol{\alpha^*}_n(\boldsymbol{\sigma}),\boldsymbol{\sigma}) ~.
\]

This is true because the projection space at step $n$ is a subspace of the one at step $n+1$. Also recall that  $E_n(\boldsymbol{\alpha^*}_n(\boldsymbol{\sigma}),\boldsymbol{\sigma})\underset{n\to \infty}{\longrightarrow} 0$, because we are dealing with finite spaces.

Second, Algorithm \ref{alg:bp} ensures that (\ref{eq:opt2}) has a unique solution, i. e. that the matrix $K \Gamma_{\boldsymbol{\sigma}}$ has full rank. Let us analyze the two possible situations:

\begin{enumerate}
\item If $\forall n \leq M$ $\max_{\boldsymbol{\sigma} \in \Sigma} E_n(\boldsymbol{\alpha_n^*}(\boldsymbol{\sigma}),\boldsymbol{\sigma}) \neq 0$, each iteration adds a support point which has a strictly positive approximation error. This means that the matrix $SH_{\sigma^*}^{-1}$ associated to this point is linearly independent from the previous basis matrices $\{SH_{\sigma_i}^{-1}\}_{i=1}^n$. Thus the updated projection basis $K\Gamma_{\sigma}$ stays full rank.

\item Conversely for $n$ such that $\max_{\boldsymbol{\sigma} \in \Sigma} E_n(\boldsymbol{\alpha_n^*}(\boldsymbol{\sigma}),\boldsymbol{\sigma}) = 0$, then all lead fields from the discretized conductivity space can be \textit{exactly} represented as a linear combination of the precomputed lead fields from support points. In practice, this naturally means that no additional support point is needed, unless a new conductivity samples are introduced.
\end{enumerate}

All these observations show that Algorithm \ref{alg:bp} guarantees that the necessary condition for Proposition \ref{prop:1} is satisfied.

\textit{Remark}: The fact that projection matrix $K \Gamma_{\boldsymbol{\sigma}}$ is full rank, and hence that matrix $\Gamma_{\boldsymbol{\sigma}}^T K^T K \Gamma_{\boldsymbol{\sigma}}$ is invertible, ensures the continuity of $\boldsymbol{\alpha}^*_n(\boldsymbol{\sigma})$ with respect to conductivity. The upper bound approximation $E_n(\boldsymbol{\alpha_n^*}(\boldsymbol{\sigma}),\boldsymbol{\sigma})$ is therefore also continuous in $\boldsymbol{\sigma}$ and hence it is bounded on any compact conductivity domain, where the exact solution exists.

Once the support points have been selected, for any new conductivity $\boldsymbol{\sigma}$, the new lead field matrix is approximated  with Algorithm \ref{alg:lf}.

\begin{algorithm}
\caption{Lead field approximation}\label{alg:lf}
\textbf{Input}: $\boldsymbol{\sigma}$; matrices $K^TK$, $K^T \vect(S)$ and $\{\widebar{L}_{ij}\}$, corresponding to support points set $\text{Supp}$.
\begin{algorithmic}[1]
\State \space\space\space\space Compute $\boldsymbol{\alpha}_n^*(\boldsymbol{\sigma})$ with (\ref{eq:opt2})
\end{algorithmic}
\textbf{Output}: $L_n(\boldsymbol{\alpha}^*_n(\boldsymbol{\sigma}),\boldsymbol{\sigma})$ with (\ref{eq:lf-approx}).
\end{algorithm}

\section{Numerical results}
\label{sec:nr}
\subsection{Data description}
To demonstrate numerically the proposed method, we use the data from \cite{Tadel}, which includes anatomical data and EEG data recorded with Yokogawa/KIT, processed with Brainstorm \cite{Tadel2011}. We use a realistic head model with three layers: brain, skull and scalp. Each surface is represented as a triangular mesh with 1082 vertices. The source space contains 15002 dipoles with fixed orientations (normal to the cortical surface).

We use the symmetric BEM implementation from OpenMEEG \cite{Kybic2005,Gramfort2010}.
 The head matrix $H$ has size $7566 \times 7566$. Matrix $D$ has size $7566 \times 15002$.

We use a model with 41 EEG electrodes, so the size of the selection matrix $S$ is $41 \times 7566$ and the lead field is a $41 \times 15002$ matrix.

Conductivity of the scalp is taken equal to 1, while the brain and skull conductivities ($\sigma_1$ and $\sigma_2$ respectively) are variable and form the parameter space (a subspace of $\mathbb{R}^2$). We are interested in a subset of this space represented by a rectangle which spans values  $(\sigma_1, \sigma_2) \in [0.5, 2] \times [10^{-4}, 10^{-1}]$. This rectangle is uniformly sampled with $25 \times 25 = 625$ points.

\subsection{Approximation error convergence}
\label{sec:conv}
We follow Algorithm \ref{alg:bp}, initializing the support point set with the four corners of the parameter space, and new support points are added one by one.

\begin{figure}[!t]

  \centering
  \subfloat[]{\includegraphics[width=0.4\textwidth]{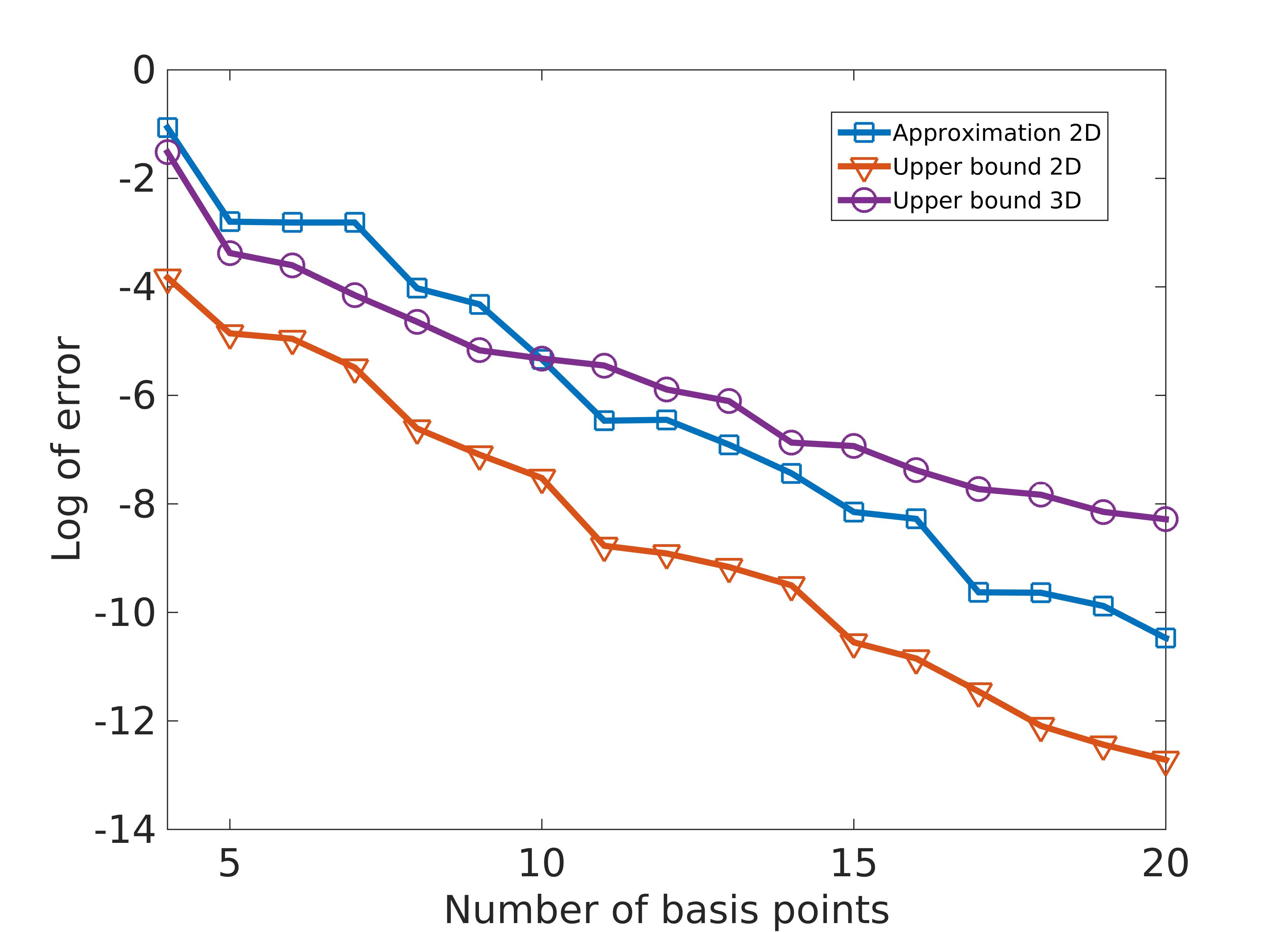}
  \label{fig:err_by_number}}
  \hfil
  \subfloat[]{\includegraphics[width=0.4\textwidth]{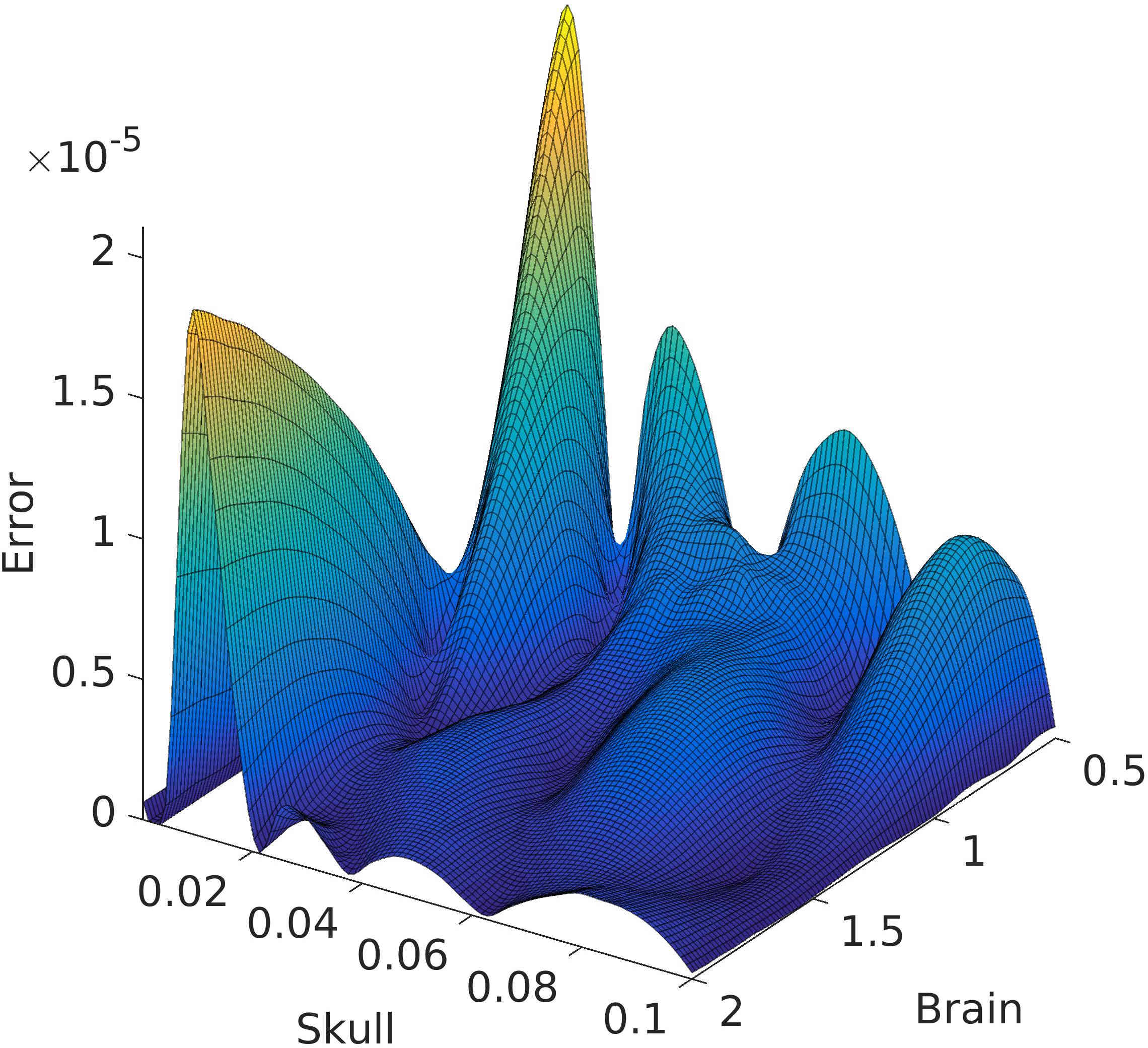}
  \label{fig:err_22bp}}
  \caption{ (a) Logarithm of maximal (over conductivity samples) upper bound and approximation errors for two and three unknown conductivities (conductivity space of dimension 2 and 3) as a function of number of support points; (b) Approximation error for 22 support points.}
  \label{error_dynamics}
\end{figure}

For each of the 625 sample points, the exact lead field is computed in order to evaluate the \textit{approximation error} (\ref{eq:error}). This is done only for validation purpose and is not necessary for the algorithm. We then compare the approximation error (\ref{eq:error}) in Fig. \ref{fig:err_by_number} with its \textit{upper bound error}, computed with (\ref{eq:prop1}). Fig. \ref{fig:err_by_number} shows that both approximation and upper bound errors ("Approximation 2D" and "Upper bound 2D") decrease exponentially fast with the number of support points. Moreover, they decrease at the same speed. This is an important property because in practice the approximation error is not available, and Fig. \ref{fig:err_by_number} shows that the decrease of the upper bound error can be used as a convergence criterion for Algorithm \ref{alg:bp}.

As shown in Fig. \ref{fig:err_22bp}, with 22 support points, the approximation error (\ref{eq:error}) over the sampled conductivity space is bounded by $2 \cdot 10^{-5}$. 

We also illustrate our method for a three-dimensional conductivity region, considering the scalp conductivity as an extra parameter. "Upper bound 3D" in Fig. \ref{fig:err_by_number} shows the decrease of the upper bound error with the number of support points. We can notice that the decrease is still exponential, even if its slope is less steep.

% \textit{Remark.} Given that there are 3 conductivities in the model, having $(\boldsymbol{\sigma}_1, \boldsymbol{\sigma}_2, \boldsymbol{\sigma}_3)$ is equivalent to $(1, \boldsymbol{\sigma}_2, \boldsymbol{\sigma}_3)$ up to a constant. 

In the case of FEM, the head matrix $H$ is homogeneous with respect to conductivities. It means that up to the constant, manifolds $H_{\boldsymbol{\sigma}}^{-1}, ~\boldsymbol{\sigma} \in \mathbb{R}^2$ and $H_{\boldsymbol{\sigma}}^{-1}, ~\boldsymbol{\sigma} \in \mathbb{R}^3$ are the same. For symmetric BEM, however, $H$ is not homogeneous (even though the lead field is) and so considering scalp conductivity as unknown does increase the dimension of the manifold $H_{\boldsymbol{\sigma}}^{-1}$.

We also evaluated the upper bound error on some conductivity points which did not correspond to any of the samples used in Algorithm \ref{alg:bp}, and it was not significantly bigger than the ones on sample points. This shows the good continuity properties of our method, mentioned previously.

\subsection{Approximation time}

The time required for computing 625 exact lead field matrices is \textit{4 hours}, which amounts to \textit{23 seconds per matrix} (2 physical cores @ 2.60GHz, 16Gb RAM). 

Precomputing all required matrices on \textit{20 support points} for Algorithm \ref{alg:lf} is achieved in \textit{14 minutes}.

Once these matrices are computed it only takes \textit{58 seconds} to approximate the 625 lead fields, which amounts to \textit{0.09 seconds per matrix}.

As mentioned in the previous section, approximation time does not depend on the complexity of the head model as it only requires to solve a least-squares problem (\ref{eq:opt2}) on the number of support points. For more complex head models, precomputation time would increase, but the approximation time would remain equal to 0.09 sec for 20 support points (but it may require more support points). 

\subsection{Conductivity estimation with simulated sources}

We now examine convergence properties in a realistic application of conductivity estimation.

Using the same head model as above, we simulate a dataset $\boldsymbol{y}$ which corresponds to a single dipole source for a reference conductivity (ground truth). Then, we use a simple dipole fitting approach to solve the inverse problem. For each of 625 conductivity samples in the domain, we approximate the lead field matrix and choose the source as the column of this matrix which best fits fits measurement in terms of Euclidean norm. Let $M_{\cdot, j}$ represent the $j$-th column of a matrix M. For each conductivity $\boldsymbol{\sigma}$ the data fitting error is defined as follows:

\[
R(\boldsymbol{\sigma}) = \min_{j,a} \left\lVert\boldsymbol{y} - a \cdot L(\boldsymbol{\sigma})_{\cdot, j}\right\rVert_2 ~,
\]
where $L(\boldsymbol{\sigma})_{\cdot, j}$ denotes $j$-th source's lead field, and $a$ - its optimal amplitude. 
$R(\boldsymbol{\sigma})$ represents the best fitting error of a single dipole corresponding to the measurement $\boldsymbol{y}$ and conductivity $\boldsymbol{\sigma}$.

Computing $R(\boldsymbol{\sigma_i})$ for each of the 625 conductivity samples, we obtain a \textit{data fitting error map} on the conductivity domain. The estimated conductivity is the one which minimizes this data fitting error.

\begin{figure}[!t]	 
  \centering
  \subfloat[]{\includegraphics[width=2.5in]{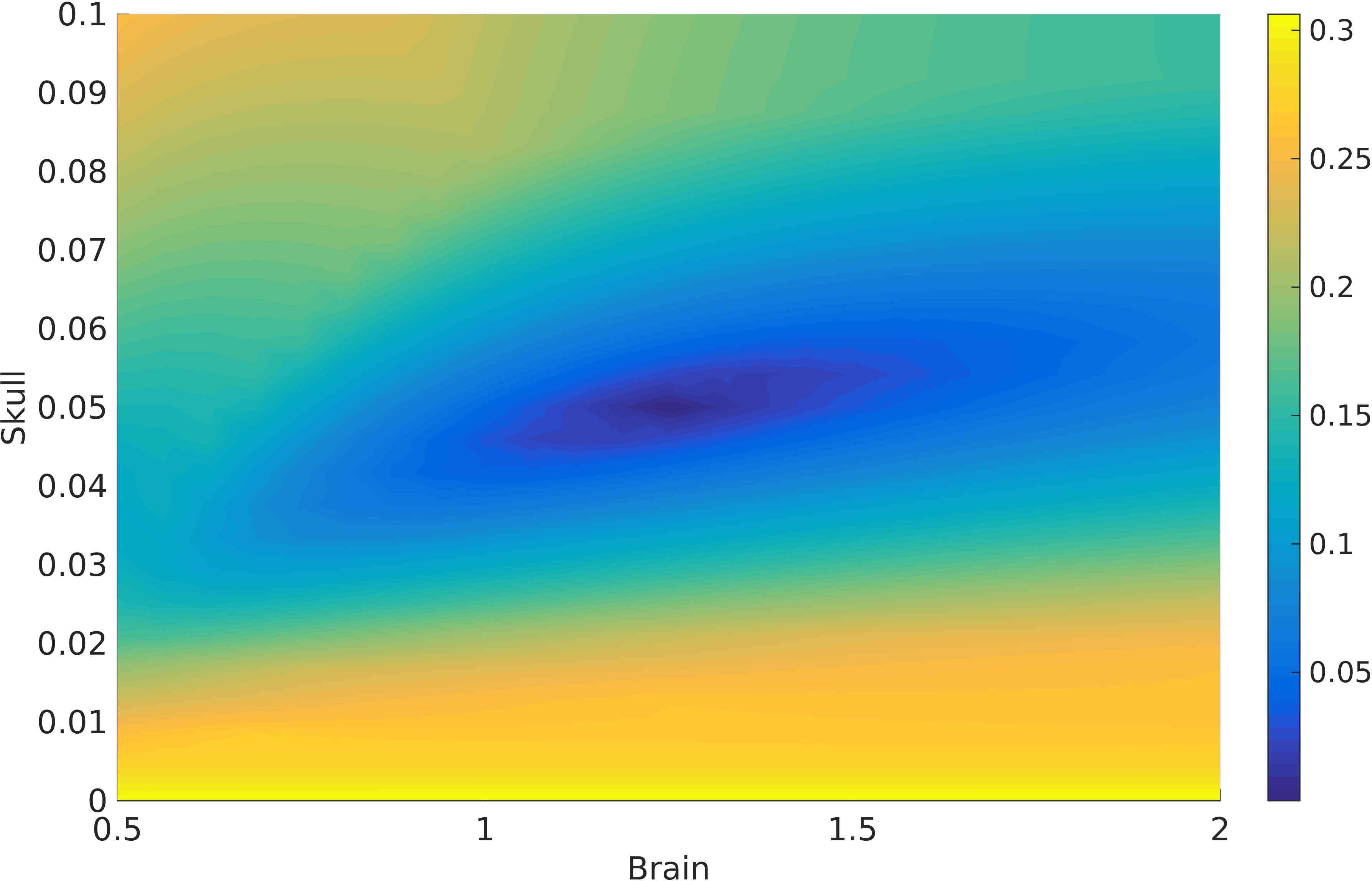}
  \label{fig:simul_data_a}} 
  \hfil
  \subfloat[]{\includegraphics[width=2.5in]{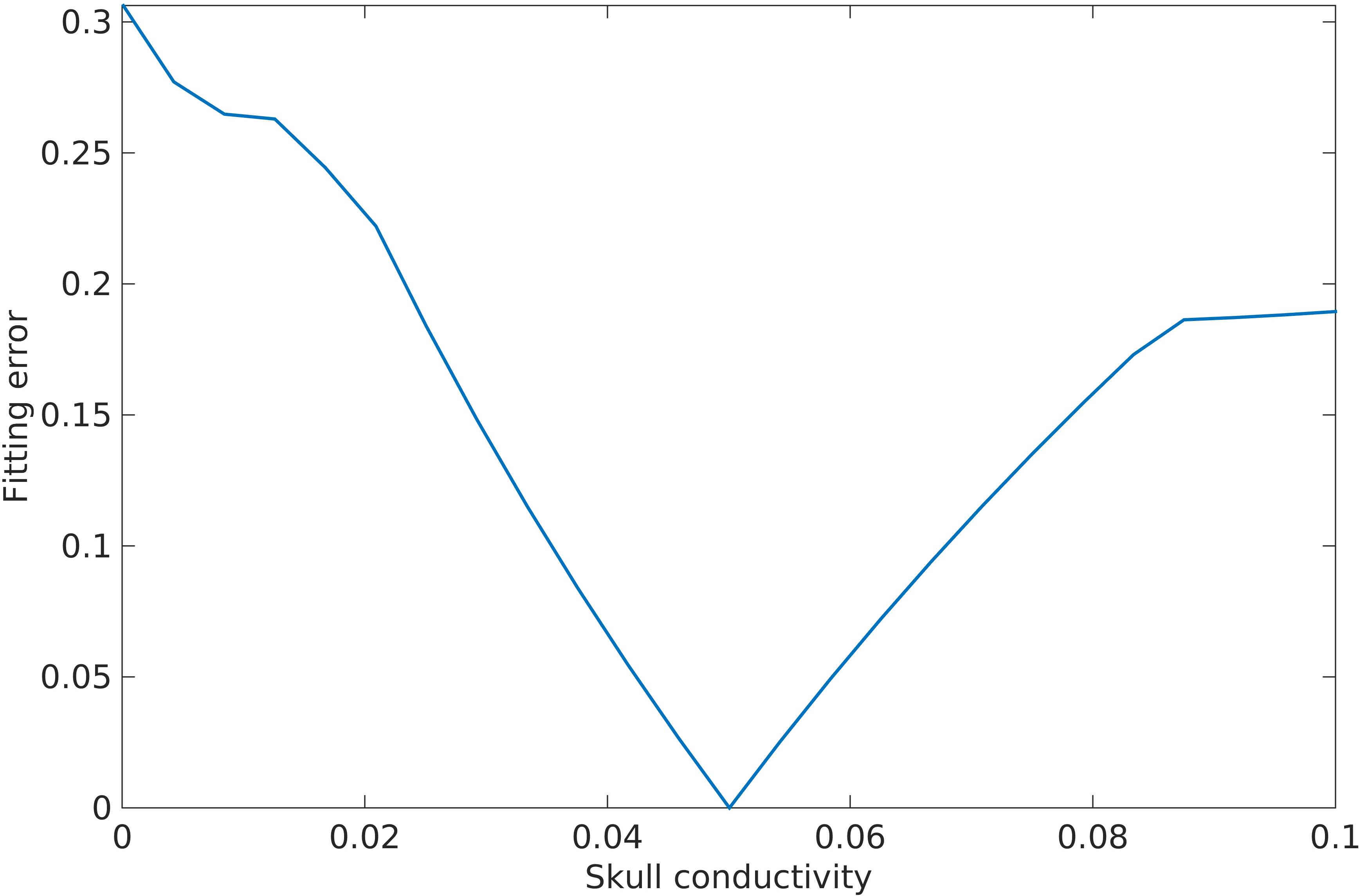}
  \label{fig:simul_data_b}}
  \caption{Data fitting error map using exact lead fields: (a) full domain of interest. The minimum of error map corresponds to simulated conductivity configuration (1.25 for brain and 0.05 for skull); (b) Fitting error with brain conductivity equal to $1.25$ (one column of map a).}
  
\end{figure}

Using the exact lead fields, we obtain a data fitting error map (Fig. \ref{fig:simul_data_a}) whose minimum lies at the simulated conductivity point (1.25 for brain and 0.05 for skull). Fig. \ref{fig:simul_data_b} shows a data fitting error for brain conductivity equal to $1.25$.

To show that approximated lead fields are of sufficient quality for conductivity estimation, we compute lead field approximation for a different number of support points. It is visible from Fig. \ref{fig:converge_simul} that the "shape" of the data fitting error map converges very fast to the exact one. \textit{10 support points} are sufficient to correctly estimate the simulated conductivity.

% \textit{Note:} For our simulation, we took the source whose lead field had  the slowest convergence of the approximation error, i. e. the "worst" source in terms of lead field approximation.

\subsection{Conductivity estimation with real data}

We use the same head model and conductivity domain of interest as in the previous section. Real EEG data was taken from a median nerve stimulation experiment. The right median nerve was percutaneously stimulated using monophasic square-wave impulses with a duration of 0.3 ms at 2.8 Hz. As in \cite{Tadel}, we filtered and averaged the data, and removed heartbeats and eye movement artifacts using Brainstorm \cite{Tadel2011}. 

%For more experiment and pre-processing details see \cite{Tadel}.

\begin{figure}[!t]
	 
  \centering
  \subfloat[]{\includegraphics[width=2.5in]{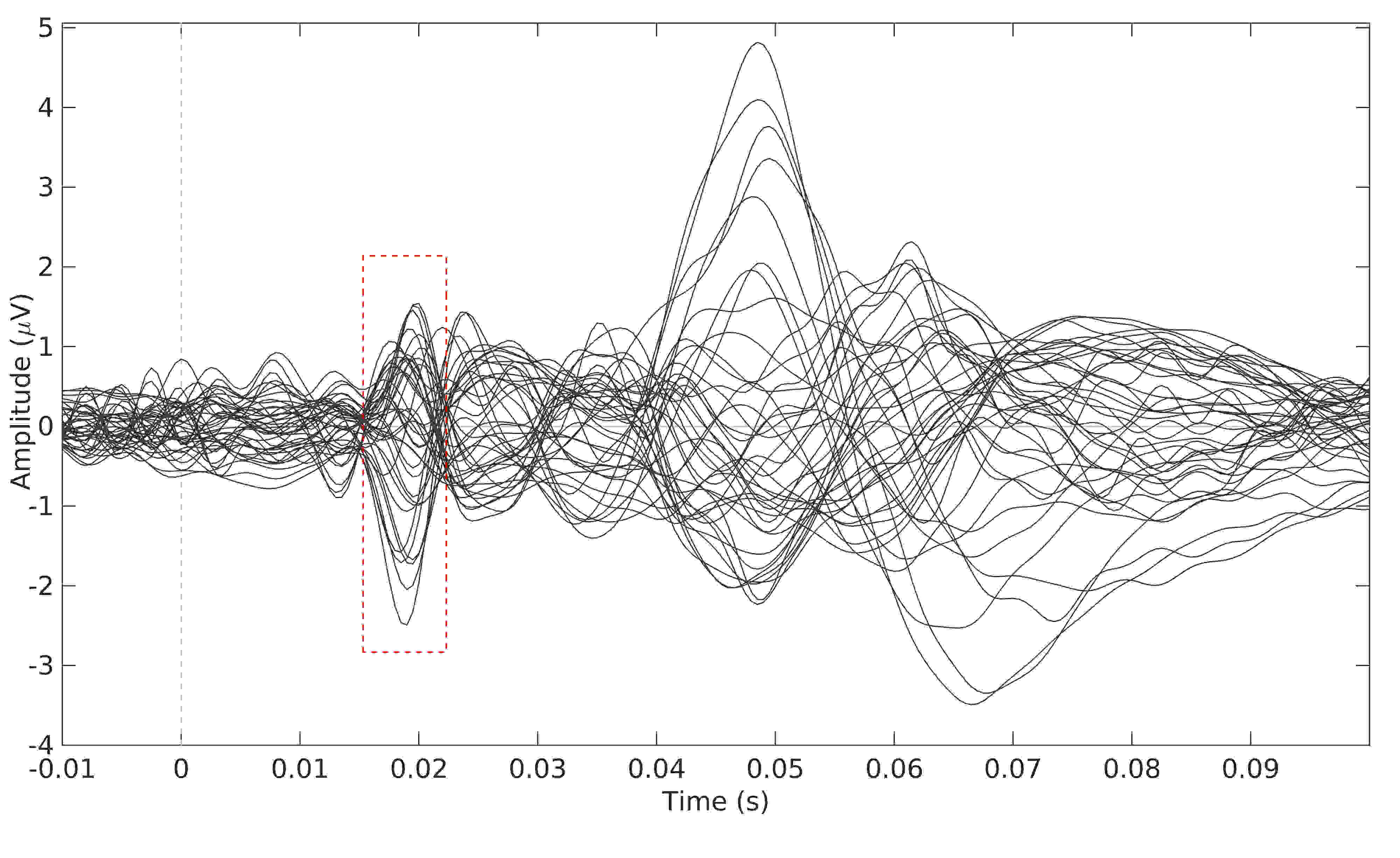}} 
  \hfil
  \subfloat[]{\includegraphics[width=2.5in]{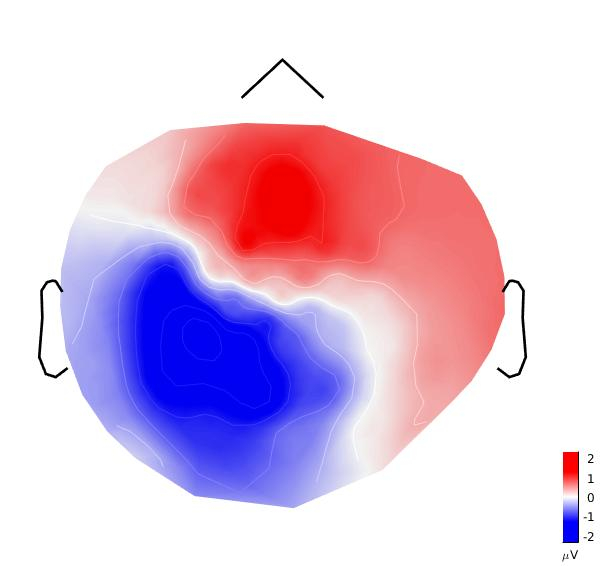}} 
  \caption{Signal of interest: (a) Averaged time series per EEG channel; (b) Topography at 19.5 ms. }
  \label{fig:real_data} 
\end{figure}

We are interested in the N20-P20 somatosensory averaged evoked potentials originating from Brodmann's area 3b situated in the posterior bank of the Rolandic fissure \cite[p.~1076]{Niedermeyer2004}.
% Electroencephalography: Basic Principles, Clinical Applications and related Fields. Ernst Niedermeyer, Fernando Lopes da Silva, p. 1076
We can see a remarkable activity peaking at approximately 20 ms which has a dipolar topography on the left hemisphere (Fig. \ref{fig:real_data}).

We analyzed 5 time samples ($\{\boldsymbol{y}(t)\}_{t=1}^5$) of a time window from 0.0185s to 0.0205s, which correspond to the local pick of the signal with dipole-like topography. The data fitting map is computed as follows:

\[
R(\boldsymbol{\sigma}) = \min_j\sum_{t=1}^5 \min_{a} \left\lVert\boldsymbol{y}(t) - a \cdot L(\boldsymbol{\sigma})_{\cdot, j}\right\rVert_2 ~.
\]

\begin{figure}[!t]	 
  \centering
  \subfloat[]{\includegraphics[width=2.5in]{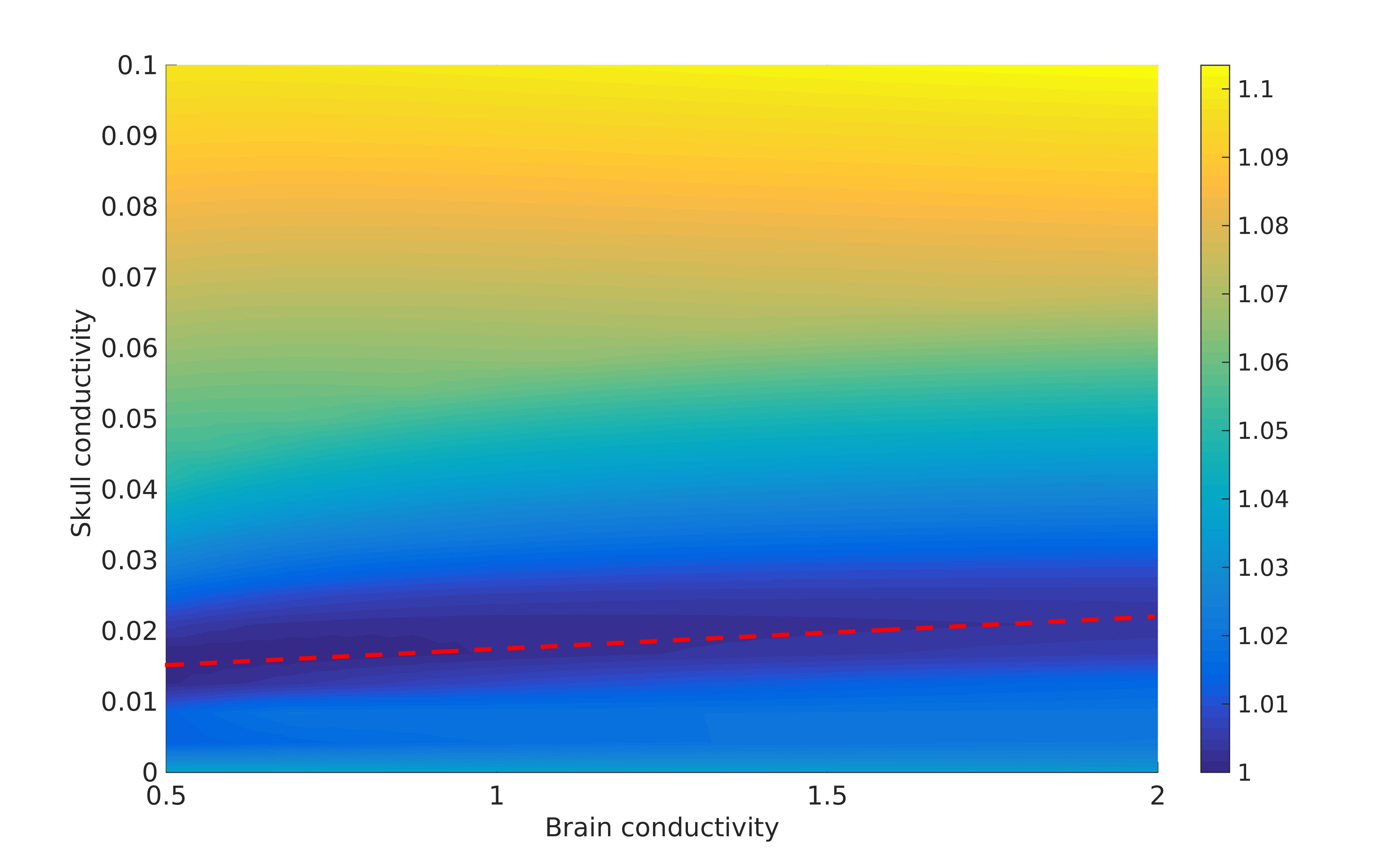}} 
  \hfil
  \subfloat[]{\includegraphics[width=2.5in]{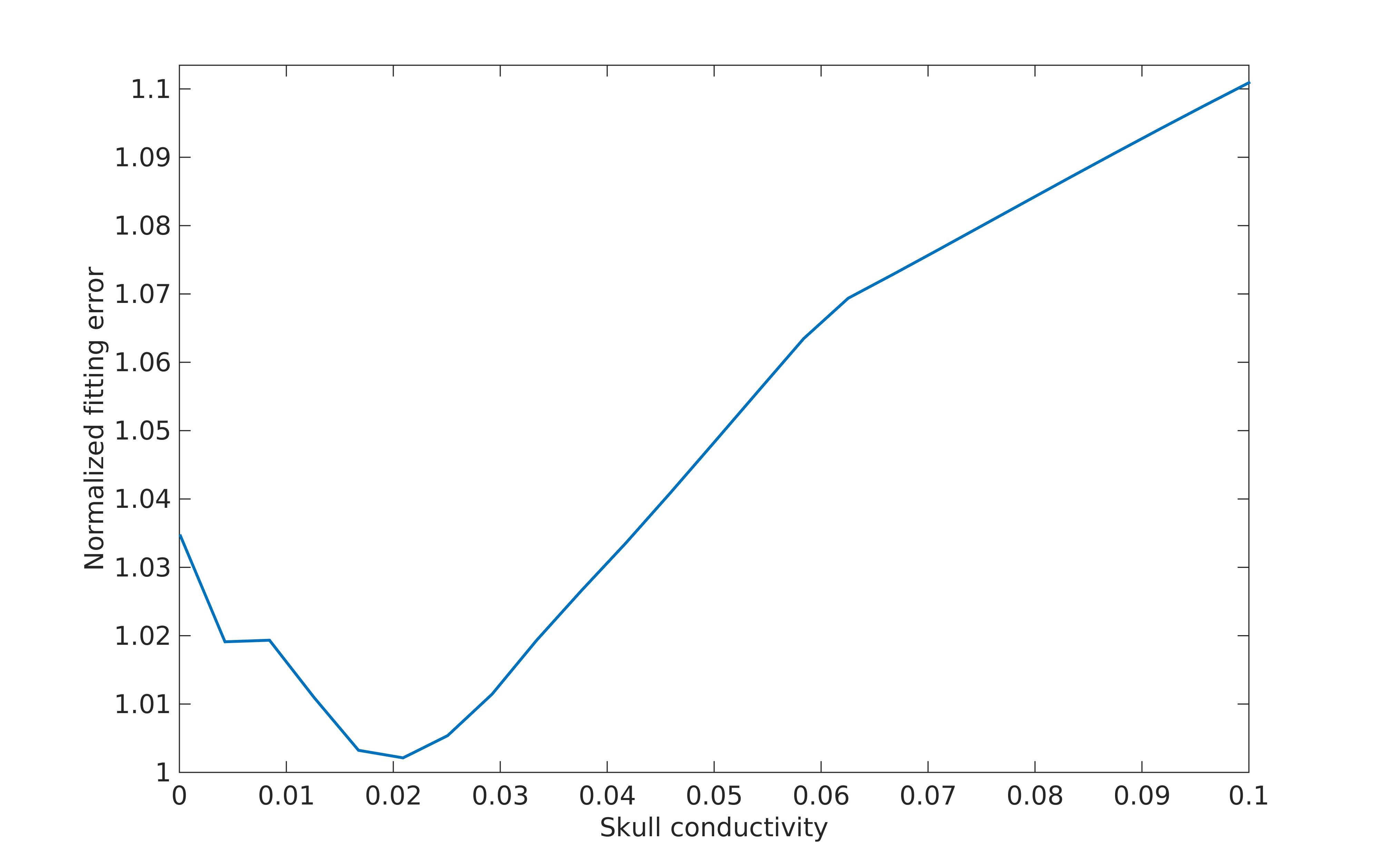}} 

  \caption{Data fitting error map using exact lead fields with real data: (a) full domain of interest. The red line represents the optimal skull conductivity for each brain conductivity; (b) Data fitting error with brain conductivity equal to $1$.}
  \label{fig:err_map_real} 
\end{figure}

The data fitting error, computed using the exact lead fields, is shown in Fig. \ref{fig:err_map_real}. It is normalized to its minimum value, i.e. $\forall i : ~R(\boldsymbol{\sigma_i}) = \frac{R(\boldsymbol{\sigma_i})}{\min_k R(\boldsymbol{\sigma_k})}$. Many factors contribute to the fitting error: additive noise, wrong conductivity model, simplified head and source models, inverse problem assumption, etc.

Because of the different sources of noise, the impact of brain conductivity becomes less important and we cannot significantly find its optimal value. But, for each fixed brain conductivity, we can obtain the skull conductivity which minimizes the error: this skull conductivity value lies between 0.01 S/m and 0.02 S/m. Let us notice that this range is clearly reduced with respect to the range reported in literature. We can also notice that conductivity contributes up to 10\% relatively to the other sources of error, which shows the importance of the correct estimation of this parameter. 
\begin{figure*}[!t]	 
  \centering
  \subfloat[]{\includegraphics[width=0.9\textwidth]{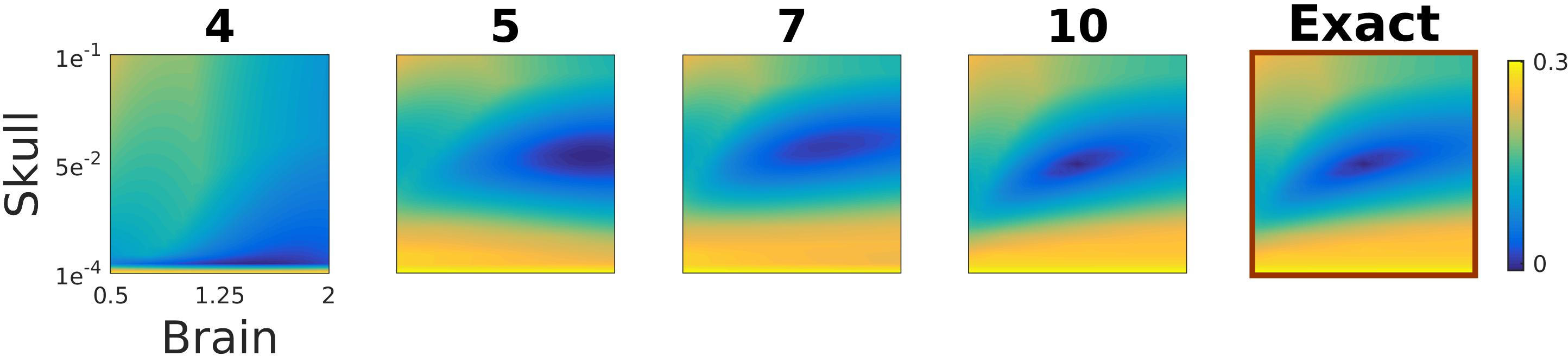}
  \label{fig:converge_simul}} 
  \hfil
  \subfloat[]{\includegraphics[width=0.9\textwidth]{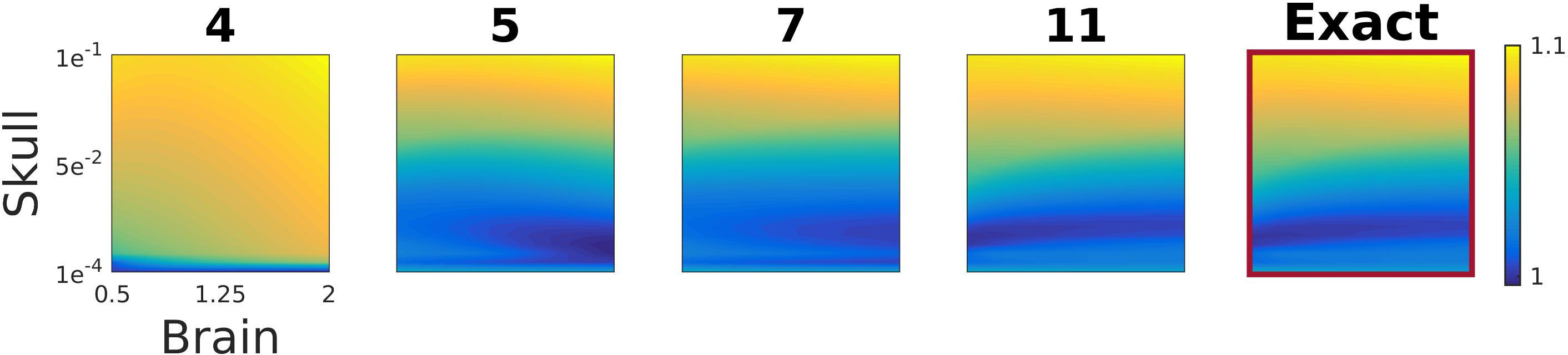}
  \label{fig:converge_real}} 
  \caption{Data fitting error map using approximated vs. exact lead fields: (a) for simulated data; (b) real data. Number on the top of the maps indicates the number of support points. For simulated and real data 10 and 11 points are respectively sufficient to obtain an error map similar to the exact one.}
  \label{fig:converge} 
\end{figure*}

As can be seen in Fig.~\ref{fig:converge_real}, it is enough to use 11 support points to obtain an error map similar to the one obtained using exact lead fields.

\section{Discussion}
Similarly to reduced basis methods used for parametrized PDEs \cite{Hesthaven2016}, our method is based on the assumption that the manifold of parametrized solutions of EEG forward problem can be approximated by low-dimensional linear subspace. As a result, we use similar tools as the reduced basis method, such as support points, upper bound error and greedy basis selection. Our approach, however, is adapted for the particular structure of the EEG forward problem. For instance, we approximate not a single solution of (\ref{eq.maxwell}), but a family of such solutions, assembled in a so-called lead field matrix: instead of dealing with a reduced basis for Galerkin projections, we deal with the projection of the inverse head matrix onto a precomputed basis.

A different approach of lead field matrix approximation is presented in~\cite{Costa2017, costa2016eeg}, where a polynomial matrix of degree $n$ is used to approximate the exact lead fields. This method has significant differences compared to ours. Firstly, this approach has essentially been developed and tested with lead fields parameterized by one single conductivity value. While it is possible to extend it to more conductivity variables, the underlying polynomial approximation problem becomes more difficult. Already with this mono dimensional approach, the choice of support points, their number and distribution is completely arbitrary with this polynomial approximation approach. Comparatively, our algorithm works with multiple conductivity dimensions and includes a basis selection, which is made in an optimal way. Furthermore, the polynomial method does not allow to control the approximation error, while ours offers some guarantees. In Appendix~\ref{app:c}, we provide a numerical comparison in 1D of our method with this polynomial approximation approach. It is shown that for a single unknown skull conductivity, the approximation error decreases faster with respect to to the number of support points with our method.  

On the other hand, the approach in~\cite{Costa2017, costa2016eeg} has several advantages. It allows to use forward solution computation as a black box without any knowledge of its structure (i. e. matrices $S$, $H_{\boldsymbol{\sigma}}$ and $D_{\boldsymbol{\sigma}}$) or of their dependence on conductivity. The approximated lead fields can be expressed in a closed form as polynomial functions of conductivity, which is easier to work with analytically (e.g. take a gradient) and can also be important for some applications.

\section{Conclusion}

In this work, we introduced a method for fast approximation of the EEG forward solution for different conductivities. Computing the exact lead field for many conductivity configurations is time consuming. Our approach only requires to compute a small number of exact solutions. Under some reasoble assumptions, all other lead fields can be approximated in a fast way with a controlled approximation error. This accelerated computing time allows to explore conductivity space, which is crucial for applications such as head tissue conductivity estimation.

Our method provides both the lead field approximation 
based on a set of given support points as well as the way to choose this set. The proposed algorithm guarantees the monotonic convergence of the approximation error. Moreover, while the complexity of computing the exact solutions for each support point depends on the number of vertices in a head model, once basis matrices are computed, the complexity of our approximation method is independent of the number of vertices.

Besides the theoretical properties of our method, we studied its empirical performance. Realistic simulations showed an exponential decrease of approximation error  with respect to the number of support points. We also demonstrated the usefulness of the method in a realistic context of conductivity estimation from EEG data, both simulated and real. As expected, a relatively small number of precomputed basis matrices provide results which are similar but remarkably faster when compared to using exact matrices.

The main motivation of this work is to propose a tool to boost lead field computation, a necessary step for simultaneous estimation of brain sources' activities and head tissues conductivity. We presented a simple approach for solving this inverse problem, based on single dipole data fitting, to show that our method can be efficiently used for this kind of problem. It would certainly be relevant to consider conductivity estimation with more complex source models (e.g. multiple dipoles), more unknown conductivities (e.g. composite skull structure \cite{Dannhauer2011}), or different conductivity space sampling approaches. We believe that our method will significantly improve the practical aspects of any such study.

\section*{Acknowledgment}
This work was supported by ANR grant VIBRATIONS (ANR-13-PRTS-0011) and by the European Research Council (ERC) under the European Union's Horizon 2020 research and innovation program (ERC Advanced Grant agreement No 694665 : CoBCoM - Computational Brain Connectivity Mapping).

\ifCLASSOPTIONcaptionsoff
  \newpage
\fi

\bibliographystyle{IEEEtran}     
\bibliography{bibl.bib}

\appendix
\subsection{Proof of Proposition \ref{prop:1}.}
\label{app:a}
\begin{IEEEproof}
\begin{align*}
&\left\lVert L(\boldsymbol{\sigma})-L_n(\boldsymbol{\alpha}, \boldsymbol{\sigma})\right\rVert_F  = \\ 
&\left\lVert  SH_{\boldsymbol{\sigma}}^{-1}D_{\boldsymbol{\sigma}} - \sum_i\alpha_i SH_{\boldsymbol{\sigma}_i}^{-1}D_{\boldsymbol{\sigma}} \right\rVert_F =\\
&\left\lVert  \Big(S - \sum_i\alpha_i SH_{\boldsymbol{\sigma}_i}^{-1}H_{\boldsymbol{\sigma}}\Big)H_{\boldsymbol{\sigma}}^{-1}D_{\boldsymbol{\sigma}} \right\rVert_F \leq \\ &\left\lVert S - \sum_i\alpha_i SH_{\boldsymbol{\sigma}_i}^{-1}H_{\boldsymbol{\sigma}}\right\rVert_F \left\lVert H_{\boldsymbol{\sigma}}^{-1}D_{\boldsymbol{\sigma}} \right\rVert_F = \\
&\left\lVert H_{\boldsymbol{\sigma}}^{-1}D_{\boldsymbol{\sigma}} \right\rVert_F E_n(\boldsymbol{\alpha},\boldsymbol{\sigma}) ~.
\end{align*}

Now we can see that $E_n$ is an upper bound of error (\ref{eq:error}):
\begin{multline*}
E(L,L_n) = \frac{\left\lVert L - L_n \right\rVert_F}{\left\lVert L \right\rVert_F} \leq \frac{\left\lVert H_{\boldsymbol{\sigma}}^{-1}D_{\boldsymbol{\sigma}} \right\rVert_F}{\left\lVert L \right\rVert_F} E_n(\boldsymbol{\alpha},\boldsymbol{\sigma}) =\\
C \cdot E_n(\boldsymbol{\alpha},\boldsymbol{\sigma}) ~.
\end{multline*}

Let $\vect(M)$ represent the  vectorization of matrix $M$. Using the fact that $\left\lVert M \right\rVert_F =\left\lVert \vect(M) \right\rVert_2$, problem (\ref{eq:prop1}) can be reformulated as:
\begin{multline*}
\boldsymbol{\alpha_n^*}(\boldsymbol{\sigma}) = \argmin_{\boldsymbol{\alpha}}\left\lVert S - \sum_i \alpha_i SH_{\boldsymbol{\sigma}_i}^{-1}H_{\boldsymbol{\sigma}}\right\rVert_F =\\ 
\argmin_{\boldsymbol{\alpha}}\left\lVert \vect(S) - \sum_i \alpha_i v(SH_{\boldsymbol{\sigma}_i}^{-1}H_{\boldsymbol{\sigma}})\right\rVert_2 ~,
\end{multline*}
which is a linear projection problem of $\vect(S)$ onto the space spanned by the set $S_n = \{ \vect(SH_{\boldsymbol{\sigma}_i}^{-1}H_{\boldsymbol{\sigma}}) \}_{i=1}^n$. It means that if $S_n$ is made of linearly independent vectors for $n_{max} = dim(\vect(S))$, then $E_{n_{max}}(\boldsymbol{\alpha}^*_{n_{max}}(\boldsymbol{\sigma}),\boldsymbol{\sigma}) = 0$. This directly implies that $E_n(\boldsymbol{\alpha^*}_n(\boldsymbol{\sigma}),\boldsymbol{\sigma})\underset{n\to \infty}{\longrightarrow} 0$ so both conditions of Definition \ref{def:opt} are verified if $S_n$ is a linearly independent set.  

\end{IEEEproof}

\subsection{Solution to problem (\ref{eq:prop1}).}
\label{app:b}
Using the decomposition of matrix $H_{\boldsymbol{\sigma}}$ (\ref{eq:sum_h}), we get:
\begin{multline*}
\sum_{i=1}^{n}\alpha_i SH_{\boldsymbol{\sigma}_i}^{-1}H_{\boldsymbol{\sigma}} = \sum_{i=1}^{n}\alpha_i SH_{\boldsymbol{\sigma}_i}^{-1}\Big(\sum_{j=1}^{N_H} \gamma_j(\boldsymbol{\sigma}) \widebar{H}_j\Big) = \\
\sum_{i=1}^{n} \alpha_i \sum_{j=1}^{N_H}  \gamma_j(\boldsymbol{\sigma}) SH_{\boldsymbol{\sigma}_i}^{-1}\widebar{H}_j ~.
\end{multline*}

Thus (\ref{eq:prop1}) becomes:
\begin{align*}
&\boldsymbol{\alpha_n^*}(\boldsymbol{\sigma}) = \argmin_{\boldsymbol{\alpha}}\left\lVert S - \sum_i^n \alpha_i SH_{\boldsymbol{\sigma}_i}^{-1}H_{\boldsymbol{\sigma}}\right\rVert_F = \\
&\argmin_{\boldsymbol{\alpha}}\left\lVert \vect(S) - \sum_{i=1}^{n} \alpha_i \sum_{j=1}^{N_H}  \gamma_j(\boldsymbol{\sigma}) \vect(SH_{\boldsymbol{\sigma}_i}^{-1}\widebar{H}_j)\right\rVert_2 = \\
&\argmin_{\boldsymbol{\alpha}}\left\lVert \vect(S) - \sum_{i=1}^{n} \alpha_i \sum_{j=1}^{N_H}  \gamma_j(\boldsymbol{\sigma}) K_{., (i-1)N_H + j}\right\rVert_2 =\\ &\argmin_{\boldsymbol{\alpha}}\left\lVert \vect(S) - K \Gamma_{\boldsymbol{\sigma}} ~ \boldsymbol{\alpha} \right\rVert_2 ~.
\end{align*}

This is the projection problem of $\vect{S}$ onto the space spanned by the columns of $K \Gamma_{\boldsymbol{\sigma}}$. Its solution is given by (\ref{eq:opt2}).

\hfill \IEEEQEDhere

\subsection{Comparison with the method of polynomial approximation.}
\label{app:c}

\begin{figure}[!t]
	\centering
		\includegraphics[width=0.4\textwidth]{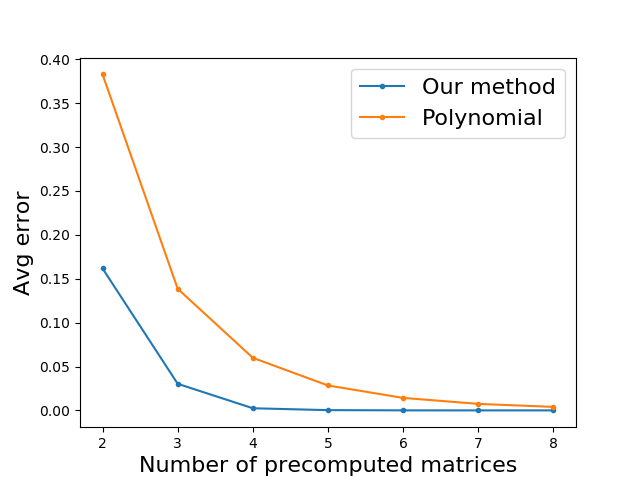}
	\caption{Relative lead field average approximation error for a 1D lead field approximation problem with respect to the number of pre-computed exact matrices, using our method and the method of polynomial approximation.}
	\label{fig:comparison}
\end{figure}

Fig.~\ref{fig:comparison} shows the comparison of our method with the polynomial approximation approach presented in~\cite{Costa2017,costa2016eeg}. In this experiment, brain and scalp conductivities are set to 1. Skull conductivity varies in the range from $10^{-4}$ to $10^{-1}$. In our method, we used a uniform mesh of 25 conductivity values, from which $n$ support points were selected using Algorithm~\ref{alg:bp}. For the polynomial approximation method, $n$ exact lead fields are computed for uniformly sampled conductivities as described in \cite{costa2016eeg}. The polynomial degree $d$ is set to $d = n-1$. The approximation error is evaluated using \textit{40 uniformly sampled exact lead fields} computed from the range of interest. The relative approximation error is computed for each of the 40 conductivities for both methods. Figure~\ref{fig:comparison} shows the average approximation error (over conductivities) with respect to the number $n$ of pre-computed lead fields (support points).    

\end{document}